\documentclass[prl,twocolumn,showpacs,amsmath,amssymb,epsfig]{revtex4-1} 
 
\usepackage{graphicx}

\begin{document}

\title{New evidence supporting the existence of the hypothetic X17 particle}

\author{A.J. Krasznahorkay}
\email{kraszna@atomki.hu}
\author{M. Csatl\'os}
\author{L. Csige}
\author{J. Guly\'as}
\author{M. Koszta}
\author{B. Szihalmi}
\author{J. Tim\'ar}
\affiliation{Institute of Nuclear Research  (Atomki),
  P.O. Box 51, H-4001 Debrecen, Hungary} 

\author{D.S. Firak}
\author{\'A. Nagy}
\author{N.J. Sas}
\affiliation{University of Debrecen, 4010 Debrecen, PO Box 105, Hungary}

\author{A. Krasznahorkay} \affiliation{CERN, Geneva, Switzerland}
\affiliation{Institute of Nuclear Research,  (Atomki),
  P.O. Box 51, H-4001 Debrecen, Hungary}

\begin{abstract}
We observed electron-positron pairs from the electro-magnetically
forbidden M0 transition depopulating the 21.01 MeV 0$^-$ state in
$^4$He.  A peak was observed in their $e^+e^-$ angular
correlations at 115$^\circ$ with 7.2$\sigma$ significance, and could be described by assuming 
the creation and subsequent decay of a light particle with mass of
$m_\mathrm{X}c^2$=16.84$\pm0.16 (stat) \pm 0.20 (syst)$ MeV and
$\Gamma_\mathrm{X}$= $3.9\times 10^{-5}$ eV.  According to the
mass, it is likely the same X17 particle, which we recently suggested 
[Phys. Rev. Lett. 116, 052501 (2016)] for describing the anomaly
observed in $^8$Be.
\end{abstract}


\maketitle

Recently, we measured electron-positron angular correlations for the
17.6 MeV, and 18.15 MeV, J$^\pi = 1^+ \rightarrow J^\pi = 0^+$, M1
transitions in $^8$Be and anomalous angular correlation was observed
\cite{kr16}.  Significant peak-like enhancement of the internal pair
creation was observed at large angles in the angular correlation of
the 18.15 MeV transition \cite{kr16}. This was interpreted as the
creation and decay of an intermediate particle X17 with mass
$m_\mathrm{X}c^2$=16.70$\pm0.35 $(stat ) $\pm 0.5 $(sys) MeV.  The possible
relation of the X boson to the dark matter problem and the fact that
it might explain the (g-2)$_\mu$ puzzle, triggered an enhanced
theoretical and experimental interest in the particle and hadron
physics community \cite{da19,ins}. 

Zhang and Miller  \cite{zh17} investigated the possibility to explain
the anomaly within nuclear physics. They explored the nuclear transition 
form factor as a possible origin of the anomaly, 
and find the required form factor to be unrealistic for the $^8$Be nucleus.

The data were explained by Feng and co-workers \cite{fe16,fe17} with a
16.7 MeV, J$^\pi$ = 1$^+$ vector gauge boson X17, which may mediate a
fifth fundamental force with some coupling to Standard Model(SM)
particles.  The X17 boson is thus produced in the decay of an excited
state to the ground state, $^8$Be$^*\rightarrow^8$Be + X17, and then
decays through the X17 $\rightarrow$ $e^+e^-$ process.

Constraints on such a new particle, notably from searches for
$\pi_0\rightarrow Z' + \gamma$ by the NA48/2 experiment \cite{ba15},
require the couplings of the $Z'$ to up and down quarks to be
‘protophobic’, i.e., the charges $e\epsilon_u$ and $e\epsilon_d$
of up and down quarks, written as multiples of the positron charge e,
 satisfy the relation $2\epsilon_u + \epsilon_d \leq 10^{-3} $
\cite{fe16,fe17}.  Subsequently, many studies of such models
have been performed including an extended two Higgs doublet model
\cite{de17}.

At the same time, Ellwanger and Moretti made another possible
explanation of the experimental results through a light pseudoscalar
particle \cite{ell16}. Given the quantum-numbers of the $^8$Be$^*$ and
$^8$Be states, the X17 boson could indeed be a $J^\pi = 0^-$
pseudoscalar particle, if it was emitted with L = 1 orbital
momentum. They predicted about ten times smaller branching ratio in
case of the 17.6 MeV transition compared to the 18.15 MeV one, which
is in nice agreement with our results.

The QCD axion is one of the most compelling solutions to the strong CP
problem. There are major current efforts in searching for an
ultra-light, invisible axion, but visible axions with decay constants
at or below the electroweak scale are believed to have been long
excluded by laboratory searches. Considering the significance of the
axion solution to the strong CP problem, Alves and Weiner \cite{al18}
revisited experimental constraints on QCD axions in the O(10 MeV) mass
window.  In particular, they found a variant axion model that remains
compatible with existing constraints. This model predicts new
particles at the GeV scale coupled hadronically, and a variety of
low-energy axion signatures, including nuclear de-excitations via
axion emission. This reopens the possibility of solving the strong CP
problem at the GeV scale.  Such axions or axion like particles (ALPs)
are expected to decay predominantly also by the emission of $e^+e^-$
pairs.

Delle Rose and co-workers \cite{de19} showed  that the anomaly can
be described with a very light Z$_0$ bosonic state, stemming from the
U(1)0 symmetry breaking, with significant axial couplings so as to
evade a variety of low scale experimental constraints.  They also
showed \cite{ro19} how both spin-0 and 1 solutions are possible and
describe the Beyond the Standard Model (BSM) scenarios that can accommodate
these. They include BSM frameworks with either an enlarged Higgs, or
gauge sector, or both.

In parallel to these recent theoretical studies, we re-investigated
the $^8$Be anomaly with an improved setup. We have confirmed the
signal of the assumed X17 particle and constrained its mass ($m_\mathrm{X}c^2 =
17.01(16)$ MeV) and branching ratio compared to the $\gamma$-decay
($B_x = 6(1)\times 10^{-6}$) \cite{kra17,kra19}.  We also  re-investigated
 the $e^+e^-$ pair correlation in the 17.6 MeV transition of
$^8$Be, in which a much smaller deviation was observed \cite{kr17}.

%%%%%%%%%%%%%%%%%%%%%%%%%%%%%%%%%%%%%%%
In order to confirm the existence of the X17 particle we have conducted
a search for its creation and decay in the 21.01 MeV  $0^-\rightarrow 0^+$  transition
of  $^4$He. Emission of a $m_\mathrm{X}c^2$ = 17 MeV vector boson (J$\pi$=1$^+$) or pseudoscalar particle
(J$^\pi$=0$^-$) is allowed in this transition with orbital angular momentum 1 or 0,
respectively. In this Letter we report on anomalous creation and angular 
correlation of electron-positron pairs in this transition, which is in
good agreement with the scenario of its decay by the assumed X17 particle.
%%%%%%%%%%%%%%%%%%%%%%%%%%%%%%%%%%%%%%%

The $^{3}$H(p,$\gamma$)$^{4}$He reaction at $E_p$=900 keV bombarding
energy was used to populate the wide ($\Gamma=0.84$ MeV) 0$^-$ second excited
state in $^4$He \cite{ti18}, located at E$_x$= 21.01 MeV. 
 This bombarding energy is below the threshold of the (p,n) reaction
(E$_{thr}$=1.018 MeV) and excites the $^4$He nucleus to E$_x$=20.49
MeV, which is  below the centroid of the wide 0$^-$ state.  A proton beam
with a typical current of 1.0 $\mu$A was impinged on a $^3$H
target. The $^3$H was absorbed in a 3 mg/cm$^2$ thick Ti layer
evaporated onto a 0.4 mm thick Mo disc.  The density of the $^3$H
atoms was $2.66\times10^{20}$ atoms/cm$^2$.  The disk was cooled down
to liquid N$_2$ temperature to prevent $^3$H evaporation.

The investigated 0$^-$ state overlaps with the first excited state in
$^{4}$He (J$^\pi$=0$^+$, E$_x$=20.21 MeV, $\Gamma$=0.50 MeV), which
was also excited but but give only a managable  background to the $e^+e^-$
spectra. 

 The experiment was performed at the 5~MV Van de Graaff accelerator in
 Debrecen.  Compared to our previous experiment \cite{kr16,gu16}, we
 increased the number of telescopes (from 5 to 6) and we replaced the
 gas-filled MWPC detectors to a double-sided silicon strip detector
 (DSSD) array.

The $e^+e^-$ pairs were detected by six plastic scintillator + DSSD
detector telescopes placed perpendicularly to the beam direction at
azimuthal angles of 0$^\circ$, 60$^\circ$, 120$^\circ$, 180$^\circ$,
240$^\circ$ and 300$^\circ$. The sizes of the scintillators are 
82$\times$86$\times$ 80 mm$^3$ each. The positions of the hits were
registered by the DSSDs having strip widths of 3 mm and a thickness of
500 $\mu$m.  The telescope detectors were placed around a vacuum
chamber made of a carbon fiber tube with a wall thickness of 1 mm.

$\gamma$ rays were also detected for monitoring.  A
$\epsilon_{rel}$=50\% HPGe detector was used at 25 cm from the target.

In the investigation of such rare processes, the cosmic ray background
needs to be taken into account. The background was measured for two
weeks before and after the experiment, and was subtracted out by using
the same gates and conditions as for the in-beam data.  In order to
determine the normalization factor for the subtraction, a 
gate of 25 MeV $\leq$E(sum)$\leq$ 50 MeV was used to determine the angular
correlations of the cosmic rays for both cases (in-beam and
off-beam). In this energy range, no in-beam counts were expected.  The
elimination of cosmic-ray background was then performed until all
events disappeared with this high-energy gate.

In order to reduce the cosmic-ray background, an active shield was
also installed above the $e^+e^-$ spectrometer.  It consisted of 12
pieces of 1.0 cm thick, 4.5 cm wide and 100 cm long plastic
scintillators. Half of the cosmic-ray yield could be suppressed in
this way.

In the original total energy spectrum of the $e^+e^-$ pairs determined using
all combinations of the telescopes we have got a very large
background from external pairs created by 
$\gamma$-rays coming from the direct proton capture, which has a cross
section of about 2 $\mu$b/sr at E$_p$= 900 keV \cite{ha95}.  In order to
reduce the background from the external pair creation, we have created
two total-energy spectra. One was constructed from $e^+e^-$ pairs,
which were detected by telescope pairs with relative angles of
120$^\circ$, while the other from $e^+e^-$ pairs which were detected
by telescope pairs with relative angles of 60$^\circ$. Since the
$e^+e^-$ pairs from the X17 boson are expected at around 115$^\circ$,
the first spectrum is expected to contain the majority of such events,
while the second is expected to be mainly background. To enhance the
X17 boson events, we subtracted the second spectrum after appropriate
normalization from the first spectrum.

Fig. 1 shows the resulting spectrum of the $e^+e^-$ pairs.

\begin{figure}[htb]
    \begin{center}
        {\includegraphics[scale=0.40]{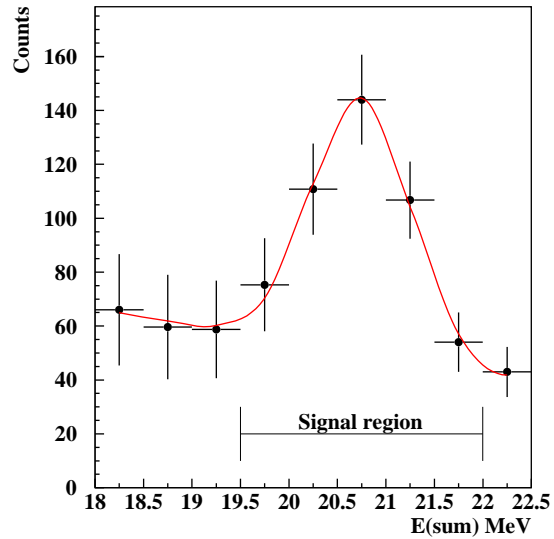}}\hspace{0.5cm}          
\caption{\it Background subtracted total energy spectrum of the $e^+e^-$ pairs. 
}
    \end{center}
\vspace{-0.5cm}
\end{figure}

The spectrum (black dots with error bars) is originated
from the 20.21 MeV E0 transition and from the  low-energy part (E$_x$=20.49 MeV) of the broad 21.01 MeV electro-magnetically
forbidden $0^- \rightarrow 0^+$ M0 transition in $^4$He.

The efficiency calibration of the telescopes was performed by using
the same dataset but with uncorrelated pairs from consecutive
events. Accordingly, an energy-independent efficiency curve could be
extracted.  The energy dependence of the efficiency calibration was
simulated by the GEANT3 code (for the same $e^+e^-$ sum-energy gate as
we used in the experimental data reduction) and taken into account as
a minor correction on the experimentally determined efficiency curve.

%A gaussian + linear background was fitted to the experimental points. 
%The result of the fit is shown as a red full curve in Fig. 1.

Fig. 2 shows our experimental results (red asterisks with error bars)
for the angular correlation of $e^+e^-$ pairs gated by the total
energy of the signal region (19.5 MeV$\leq E_{tot}\leq$22.0 MeV), using the asymmetry
parameter ($-0.5\leq y \leq 0.5$) as defined in Ref.\cite{kr16} and corrected for the relative efficiency of the spectrometer. Black
stars with error bars show the angular correlation of $e^+e^-$ pairs for the background region
(5 MeV$\leq E_{tot}\leq$19 MeV and $-0.5\leq y \leq 0.5$). 

\begin{figure}[htb]
    \begin{center}
        {\includegraphics[scale=0.45]{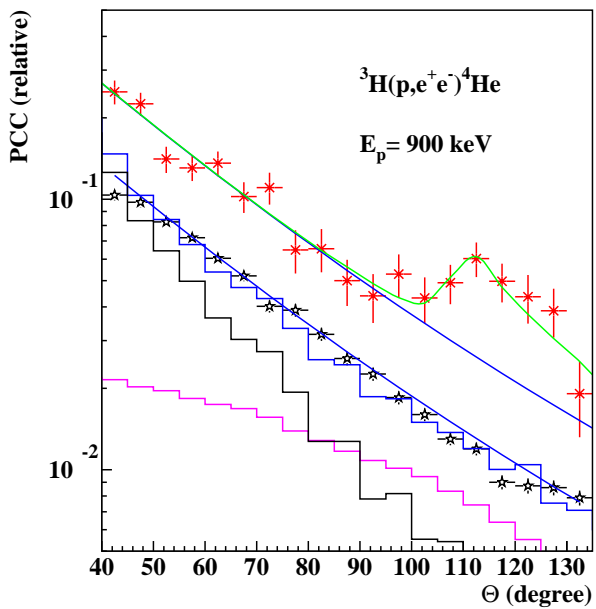}}\hspace{0.5cm}          
\caption{\it Angular correlations for the  $e^+e^-$ pairs measured in 
the $^{3}$H(p,$\gamma$)$^{4}$He reaction at the $E_p$=900 keV.}
    \end{center}
\vspace{-0.5cm}
\end{figure}

According to our simulations such background is 
 originated from external pair creation on the target backing  and on the surrounding materials 
(black histogram) and from the IPC $e^+e^-$ pairs crated  in the  
J$^\pi$=0$^+ \rightarrow$ 0$^+_{gs.} $ E0 transition (magenta histogram). The sum of that two components
fitted to the experimental data 
is shown as a blue histogram.
The data measured for the background were
fitted by a 4-th order exponential polynomial, and the result is shown in a
blue full curve. This blue curve was rescaled to fit the background of the angular correlation
shown in red in the range of $40^\circ\leq\theta\leq90^\circ$. 
The obtained experimental angular correlation exhibits a rather sharp bump
at around 115$^\circ$. This feature is similar to the anomaly observed in $^8$Be,
and seems to be in agreement with the X17 boson decay scenario.
The
green full curve shows the simulated angular correlation including the
decay of the expected X17 particle, which was fitted to the data.
In order to derive the exact value for the mass of the decaying particle from
the present data, we carried out a fitting procedure for both the mass value 
and the height of the observed peak.

The fit was performed with RooFit \cite{Verkerke:2003ir} by describing
the $e^+e^-$ angular correlation distribution with the following
probability density function (PDF):
\begin{equation}
PDF(e^+e^-) = N_{Bg} * PDF(exp) + N_{Sig} * PDF(sig)\ ,
\label{eq:pdf}
\end{equation}

\noindent
where $PDF(exp)$ was determined experimentally for the background region, $PDF(sig)$ 
was simulated by GEANT4 for the two-body decay of the X particle as a function 
of its mass, and   $N_{Bg}$ and $N_{Sig}$ are the fitted number of background and
signal events, respectively.

The signal PDF was constructed as a 2-dimensional model function
of the $e^+e^-$ opening angle and the mass of the simulated
particle. To construct the mass dependence, the PDF linearly
interpolates the $e^+e^-$ opening angle distributions simulated for
discrete particle masses.

Using the composite PDF described in Equation~\ref{eq:pdf} we first
performed a list of fits by fixing the simulated particle mass in the
signal PDF to a certain value, and letting RooFit estimate the best
values for $N_{Sig}$ and$N_{Bg}$.  Letting the particle mass lose in
the fit, the best fitted mass is calculated for the best fit and shown
also in Fig. 2.  in green.
The significance of the peak observed in the $e^+e^-$ angular
correlations was found to be 7.2$\sigma$.  
 The mass of the particle derived from the fit is: $m_\mathrm{X}c^2$=16.84$\pm0.16$ MeV. 

The partial width of the boson-decay $\Gamma_\mathrm{X}$ from the 0$^-$-state to the ground state is 
estimated as follows: 
\begin{equation}
 \Gamma_\mathrm{X}/\Gamma_{E0}= \left({\sigma(X17)\over\sigma(E0)}\right)_{exp.} 
\left({ \sigma(0^+)\over\sigma(0^-)} \right)_{th} .
\end{equation}
The  $\sigma$(X17)/$\sigma$(E0)= 0.20 was 
obtained from the fit of the simulated $e^+e^-$ angular correlations to our data.
The exciataion energy of the nucleus in the case of E$_p$=900 keV is 20.49 MeV.
At that energy the contribution of the  0$^+$ and  0$^-$ resonances are equal, so the ratio of the cross sections in the resonant proton capture 
reaction could be calculated as follows:
\begin{equation}
\left({\sigma(0^+)\over\sigma(0^-)}\right)_{th}={\Gamma_{tot}(0^+)\over\Gamma_{tot}(0^-)}
  = 0.59.
\end{equation}

Since $\Gamma_{E0}$=$(3.3\pm 1)\times 10^{-4}$ is known \cite{wa70}, than 
$\Gamma_\mathrm{X}$=$0.2\times0.59\times 3.3 \times 10^{-4} = 3.9\times 10^{-5}$ eV.

In the case of $^8$Be it was $\Gamma_\mathrm{X}$= 
$\Gamma_\gamma\times B_X= 1.9\times6\times 10^{-6}$ eV = 1.2$\times10^{-5}$ eV. 
This value is indeed expected to be smaller due to the  phase space correction factor.

The invariant mass distribution was also 
calculated from the
measured energies and angles of the same dataset:
$$m_\mathrm{X}c^2 = \sqrt{1-y^2}E \sin(\theta/2) + 2m_e^2\left(1+{(1+y^2)\over(1-y^2)}\cos(\theta)\right) , $$
where $E=E_{e^+} + E_{e^-}$ and $y=(E_{e^+} - E_{e^-})/(E_{e^+} + E_{e^-}) .$ 
The result is shown in Fig. 3  for the signal (19.5~MeV$\leq E_{tot}\leq$22.0 MeV, in red) and 
background (5 MeV$\leq E_{tot}\leq$19 MeV, in black) regions.

\begin{figure}[t]
    \begin{center}
        {\includegraphics[scale=0.4]{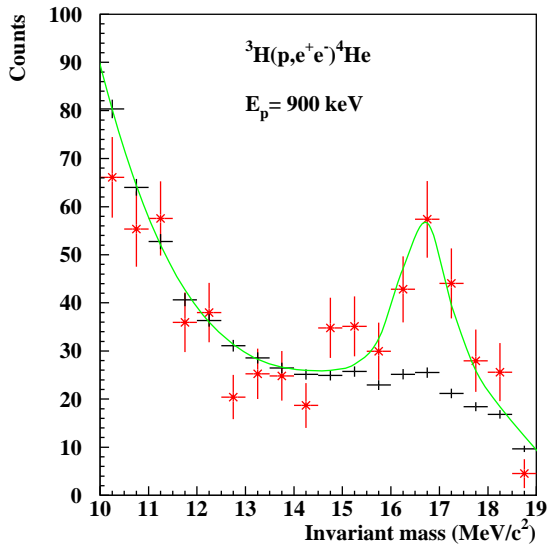}}\hspace{0.5cm}          
\caption{\it Invariant mass distribution derived for the 20.49 MeV 
transition in $^4$He.}
    \end{center}
\vspace{-0.5cm}
\end{figure}

The observed local p$_0$ probability as a function of $m_\mathrm{X}$, associated to the invariant mass
distribution is shown in Fig. 4. It is the probability that the observed excess
is due to a statistical oscillation of the background, as defined and used in
high energy physics \cite{gr17}.

\begin{figure}[htb]
    \begin{center}
        {\includegraphics[scale=0.40]{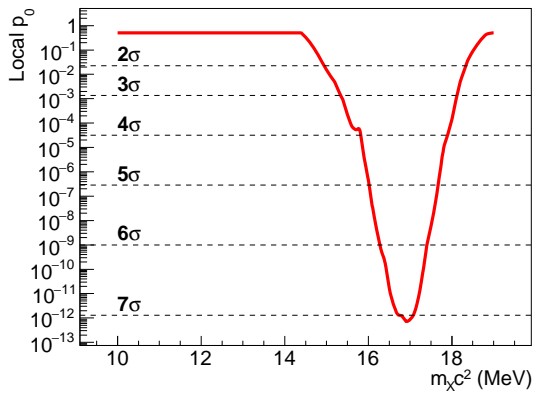}}\hspace{0.5cm}          
\caption{\it  The observed local p$_0$ as a function of the hypothesized 
X17 boson mass for the X17 $\rightarrow  e^+e^-$ channel.}
    \end{center}
\vspace{-0.5cm}
\end{figure}

The significance of the peak
observed in the $e^+e^-$ invariant mass distribution was found to be
7.1$\sigma$.  The mass of the particle derived from the fit is:
$m_\mathrm{X}c^2$=17.00$\pm0.13$ MeV. This value agrees within the erroor bar
with the one we derived from the fit of the angular correlation.

The systematic uncertainties was estimated by taking into account the
uncertainty of the target position along the beam line, which was
estimated to be $\pm$ 2 mm, which may cause $m_\mathrm{X}c^2$$\pm$ 0.06 MeV
uncertainty. The uncertainty of the place of the beam spot
perpendicular to the beam axis was estimated to be in worst case also
$\pm$ 2 mm, which may cause a shift in the invariant mass of  $m_\mathrm{X}c^2$$\pm$
0.15 MeV/c$^2$. The whole systematic error was conservatively estimated as: $m_\mathrm{X}c^2\pm$0.20
MeV.

The obtained mass value agrees very well with that of derived for the X17 
boson from the $^8$Be experiments. This is remarkable taking into account the
fact that in the present $^4$He transition the anomalous bump in the angular
correlation spectrum appears at a quite different angle than it appears in
the $^8$Be experiments due to the different energies of the two excited states.
The good agreement between the two masses leads to the scenario of decaying
both studied excited states by the same X17 particle. This strengthens the
validity of the X17 boson hypothesis.
It is also worth mentioning that strictly speaking it cannot be proved that
in the $^4$He case the anomalous decay belongs to the 21.01 MeV  0$^- \rightarrow 0^+$ 
transition. The wide 20.21 MeV 0$^+$ first excited state overlaps with the 
21.01 MeV 0$^-$ state, and they both were populated in the experiment. However,
the anomalous decay of the 0$^+$ state would result a different new particle 
than the decay of the 0$^-$ state or the decay of the 1$^+$ state in the $^8$Be.
Assuming two new particles with the same mass is a less probable scenario
than assuming only one X17 particle, which explains both anomalies.

We are expecting independent (particle physics) experimental results
to come in the coming years. In the following we cite a few of them.

Recently, the NA64 experiment \cite{ba18} at CERN presented the first
direct search with a 100 GeV/c e$^-$ beam for this hypothetical $m_\mathrm{X}c^2$=16.7
MeV boson and excluded part of its allowed parameter space, but
left the still unexplored region $4.2\times
10^{-4}\leq\epsilon_e\leq1.4\times10^{-3}$ as quite an exciting
prospect for further research.  Experiment will be continued
\cite{ki19,ba19}.

The goal of ForwArd Search ExpeRiment (FASER) \cite{fe18} at the LHC
is to discover light, weakly interacting particles with a small (1
m$^3$) detector placed in the far-forward region of ATLAS. In
particular, Ariga and his coauthors \cite{ar18, ar19, ar19a, ar19b,
  ar19c} considered the discovery prospects for ALPs.  
The project has already been approved, and the experiment
will start in 2023.

Jiang, Yang and Qiao \cite{ji19} presented a comprehensive
investigation on the possibility of search for the X boson directly in
$e^+ - e^-$ collisions, and through the decay of the created $J/\psi$
particles at the BESIII experiment for both spin-0 and spin-1
hypotheses. They suggest that Z$_0$-like boson signal might be found
or excluded in the present run of BESIII. The BESIII experiment has
accumulated the largest J/$\psi$ dataset ( 10$^{10}$ J/$\psi$ events)
worldwide. They found that this is an ideal channel to test the spin
of the particle.  They are expecting $\approx10^{3}$ scalar/Z$_0$-like
X bosons when setting the reduced Yukawa coupling parameters to
10$^{-3}$, which is within the analysis sensitivity of BESIII.

Nardi and coauthors \cite{na18} suggested the resonant production of
X17 in positron beam dump experiments. They explored the foreseeable
sensitivity of the Frascati PADME experiment in searching with this
technique for the X17 boson invoked to explain the $^8$Be anomaly in
nuclear transitions.  PADME already took  some test data and is running
until the end of 2019 \cite{ta18, ca18, pi19, ko19, sp18}. After that,
the experimental setup will be moved to Cornell and/or JLAB to get
higher intensity positron beams.

DarkLight will search for 10 - 100 MeV/c$^2$ dark photons
\cite{co17}. The sensitivity is projected to reach the $^8$Be anomaly
region. The first beam was already used  in summer 2016. Currently,
they are doing proof-of-principle measurements \cite{wa19}.

In summary, we have observed $e^+e^-$ pairs from an
electro-magnetically forbidden M0 transition depopulating the 21.01
MeV 0$^-$ state in $^{4}$He. The energy sum of the pairs corresponds
to the energy of the transition. The measured $e^+e^-$ angular
correlation for the pairs shows a peak at 115$^\circ$, supporting the
creation and decay of the X17 particle with mass of
$m_\mathrm{X}c^2$=16.84$\pm0.16 (stat) \pm 0.20 (syst)$ MeV. This mass agrees
nicely with the value of $m_\mathrm{X}c^2$=17.01 $\pm0.16$ MeV we previously 
derived in the $^8$Be experiment \cite{kr16,kra17,kra19}.  
The partial width of the X17 particle decay is esimated to be: $\Gamma_\mathrm{X}$= $3.9\times 10^{-5}$ eV.
We are expecting more, independent experimental results to come 
for the X17 particle in the coming years.

We wish to thank D. Horv\'ath for the critical reading the manuscript
and for the many useful discussions. We wish to thank also for Z. Pintye for 
the mechanical design of the experiment.
This work has been supported by the Hungarian NKFI Foundation No.\,
K124810, by the GINOP-2.3.3-15-2016-00034 and by the J\'anos Bolyai
Research Fellowship of the Hungarian Academy of Sciences (L. Csige).

\

\end{document}